\documentclass[floatfix,twocolumn,prl,longbibliography]{revtex4}
\usepackage{graphicx}
\usepackage{dcolumn}
\usepackage{bm}
\usepackage{amsmath}
\usepackage{times}
\usepackage{color}
\usepackage[breaklinks=true,colorlinks,citecolor=blue,linkcolor=blue,urlcolor=blue]{hyperref}
\usepackage{gensymb}

\makeatletter

\newcommand{\Rmnum}[1]{\expandafter\@slowromancap\romannumeral #1@}
\makeatother

\begin{document}
\title{Tuning the chiral orbital currents in a colossal magnetoresistive nodal line ferrimagnet}
\author{Arnab Das}
\affiliation{Department of Physics, Indian Institute of Technology Kanpur, Kanpur 208016, India}

\author{Soumik Mukhopadhyay}
\email{soumikm@iitk.ac.in}
\affiliation{Department of Physics, Indian Institute of Technology Kanpur, Kanpur 208016, India}  

\begin{abstract}
    The ferrimagnetic nodal-line semiconductor Mn$_3$Si$_2$Te$_6$ exhibits colossal magnetoresistance (CMR) owing to the chiral orbital currents (COC). The COC is developed due to spin-orbit interaction (SOI) attributed to the tellurium (Te) atoms. Here, we observe that on chemical substitution of the Te atoms with selenium (Se), the COC, which runs along the Te-Te edges of the MnTe$_6$ octahedra, becomes weaker and thus affects the angular magnetoresistance (MR) of Mn$_3$Si$_2$Te$_6$. We find that the application of magnetic field along the easy axis leads to a considerable drop in resistance in substituted crystals, which otherwise exhibits weak MR. On the other hand, the CMR effect along the partially polarized magnetization direction is found to be only marginally affected due to the substitution and persists even for a significantly high concentration of Se.
\end{abstract}

\maketitle

{\it Introduction:---}
The phenomena of colossal magnetoresistance (CMR) is not new and have been under study since the 1990's when it was first discovered in a hole-doped perovskite system \cite{lcmo1, lcmo2, lcmo3, lcmo4, lcmo5, lcmo6, lcmo7, lcmo8, lcmo9, lcmo10, lcmo11, lcmo12}. In such systems, the CMR is governed by the interplay of double exchange mechanism and dynamic Jahn-Teller effect which plays a key role in bringing forth the concurrent magnetic and metal-insulator transition \cite{de1, de2, de3, de4}. Later on, however, materials like Ti$_2$Mn$_2$O$_7$ \cite{ti1, ti2}, Eu$_5$In$_2$Sb$_6$ \cite{eu1}, EuCd$_2$P$_2$ \cite{eu2}, EuTe$_2$ and EuMnSb$_2$ \cite{eu3, eu4, eu5}, have been discovered that shows CMR without following the conventional mechanism typical of doped perovskites. In recent times, two-dimensional (2D) layered correlated systems with transition atoms have gained much attention due to their rich physical properties, including CMR \cite{Mndoped paper}. In layered ferrimagnetic (FIM) nodal-line semiconductor Mn$_3$Si$_2$Te$_6$, a very large CMR is exhibited which forces the resistivity to drop by several orders of magnitude. However, this sharp decrease in resistance occurs only when the magnetic field is applied along the magnetic hard axis  \cite{Seo2021, PhysRevB.95.174440}. Thus, the CMR in Mn$_3$Si$_2$Te$_6$ is not dictated by the spin polarization. Rather, it occurs only when the spin polarization is avoided \cite{PhysRevB.103.L161105}. This CMR is quite unique and shows a current dependency, decreasing with the increase in current, and finally vanishes at a particular threshold current, I$\mathrm{_C}$ \cite{Zhang2022}. The origin of this unique CMR in Mn$_3$Si$_2$Te$_6$ has been attributed to the breaking of nodal-line degeneracy. The spin-orbit coupling (SOC) between Mn and Te shifts one of the bands towards the Fermi level, inducing a metal-insulator transition when the field is along the out-of-plane (OOP) direction \cite{Seo2021,dft_paper_doping1, Susilo_paper}. An alternative perspective focuses on the chiral-orbital currents (COC) that flow along the Te edges in the ab-plane of Mn$_3$Si$_2$Te$_6$. These currents generate a moment, $\mathrm{M_{COC}}$, along the c-axis, which couples with the moments of the Mn atoms, M$\mathrm{_{Mn}}$. When the magnetic field is along the c-axis, the combined effect of $\mathrm{M_{COC}}$ and M$\mathrm{_{Mn}}$ governs the observed magnetization and CMR effect \cite{Zhang2022, Zhang2024, mag_curr,
anapole}. However, till now, most of the interesting phenomena about Mn$_3$Si$_2$Te$_6$ have been found along the c-axis only, which surprisingly is not the magnetic easy axis. There have been very few reports about the properties of Mn$_3$Si$_2$Te$_6$ along the easy plane (the ab-plane), exacerbated by the fact that in the ab-plane, the magnetoresistance (MR) is low. Thus, further research to improve the MR response along the magnetic easy plane and connecting it with the existing theories is crucial to draw a complete picture of the physics of Mn$_3$Si$_2$Te$_6$ and also for future applications.

In our work, we observe that by progressive substitution of the tellurium with selenium, the negative MR along the ab-plane increases. Additionally, this substitution has very little effect on the already high CMR observed along the c-axis. Our calculations reveal that the MR behaviour of Mn$_3$Si$_2$Te$_6$ is deeply linked with the magnetisation anisotropy and the resulting modification of COC. 

{\it Crystal Growth:---}
Single crystals of Mn$_3$Si$_2$Te$_6$ have been prepared using the chemical vapour transport (CVT) method \cite{PhysRevB.105.214405} where iodine is used as the transport agent. The initial precursors, Mn (99.95$\%$, Alfa Aesar), Si (99.999$\%$, Alfa Aesar), and Te (99.99$\%$, Alfa Aesar) powders, are mixed in the stoichiometric ratio of 3:2:6. The powders are homogeneously mixed and pelletized, and then kept in a vacuum-sealed quartz ampule along with 40 mg of I$_2$. The growth process utilizes a TG3-1200 gradient three-zone tube furnace with the hot end kept at 800\degree C and the cold end maintained at 750\degree C. Both ends are heated at a rate of 60\degree C/hr. The ampoule is kept in the furnace for 20 days, and after that, it is cooled to room temperature and taken out. The resultis the formation of crystals of 1-2 mm in size.

For substituting tellurium (Te) by selenium (Se) with different concentrations, we employ similar CVT technique. Here, Mn (99.95$\%$, Alfa Aesar), Si (99.999$\%$, Alfa Aesar), Te (99.99$\%$, Alfa Aesar) and Se (99.999$\%$, Alfa Aesar) powders, are mixed in the stoichiometric ratio. Again, the mixture is pelletized and kept in an ampoule and placed in the three-zone furnace. This time, the hot zone is kept at 750\degree C and the cold end at 700\degree C and the ampoule is kept in the furnace for 17 days. Upon cooling to room temperature, we successfully obtain the crystals.
\begin{figure}[htp]
  \includegraphics[width=1\linewidth]{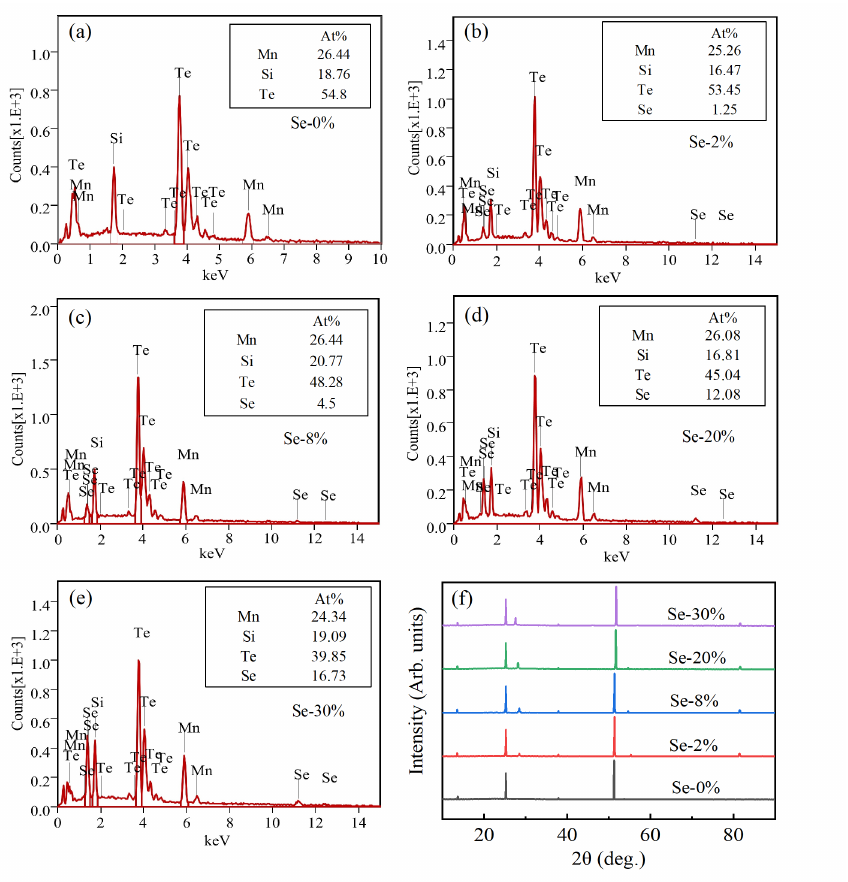}
  \caption{(a) - (e) Energy dispersive spectroscopy data of Mn$_3$Si$_2$(Te$_{1-x}$Se$_x$)$_6$ single crystals for different values of $x$. (f) XRD Data of Mn$_3$Si$_2$(Te$_{1-x}$Se$_x$)$_6$ single crystals for different values of $x$.}
  \label{fig1}
\end{figure}

{\it Characterization:---}
X-ray diffraction (XRD) and energy dispersive spectroscopy (EDS) measurements are performed to determine the crystallinity as well as the elemental composition of both the substituted and unsubstituted sample. The stoichiometry of all the samples is confirmed from the EDS measurements. Fig.-\ref{fig1}(a)-(e) presents the EDS spectra with prominent Mn, Si, Te and Se peaks, obtained from freshly cleaved surfaces of Mn$_3$Si$_2$(Te$_{1-x}$Se$_x$)$_6$ single crystals for different $x$ values. The atomic percentage is also highlighted in the inset for each grown crystal, which confirms that the obtained stoichiometry is close to the targeted one. The X-ray diffraction (XRD) measurement is carried out using a PANalytical X’Pert diffractometer to confirm the high crystallinity of the grown crystals. Fig-\ref{fig1}(f) shows the XRD data of the as grown single crystals for different $x$ values, which reveal the (001) orientation of the crystal surface. The temperature dependence of magnetization and the isothermal magnetization hysteresis measurements have been carried out using a Quantum Design physical properties measurement system (PPMS). The electrical transport properties have been measured in a CRYOGENIC variable temperature insert (VTI) cryostat using the standard lock-in technique.
\begin{figure}[]
  \includegraphics[width=0.8\linewidth]{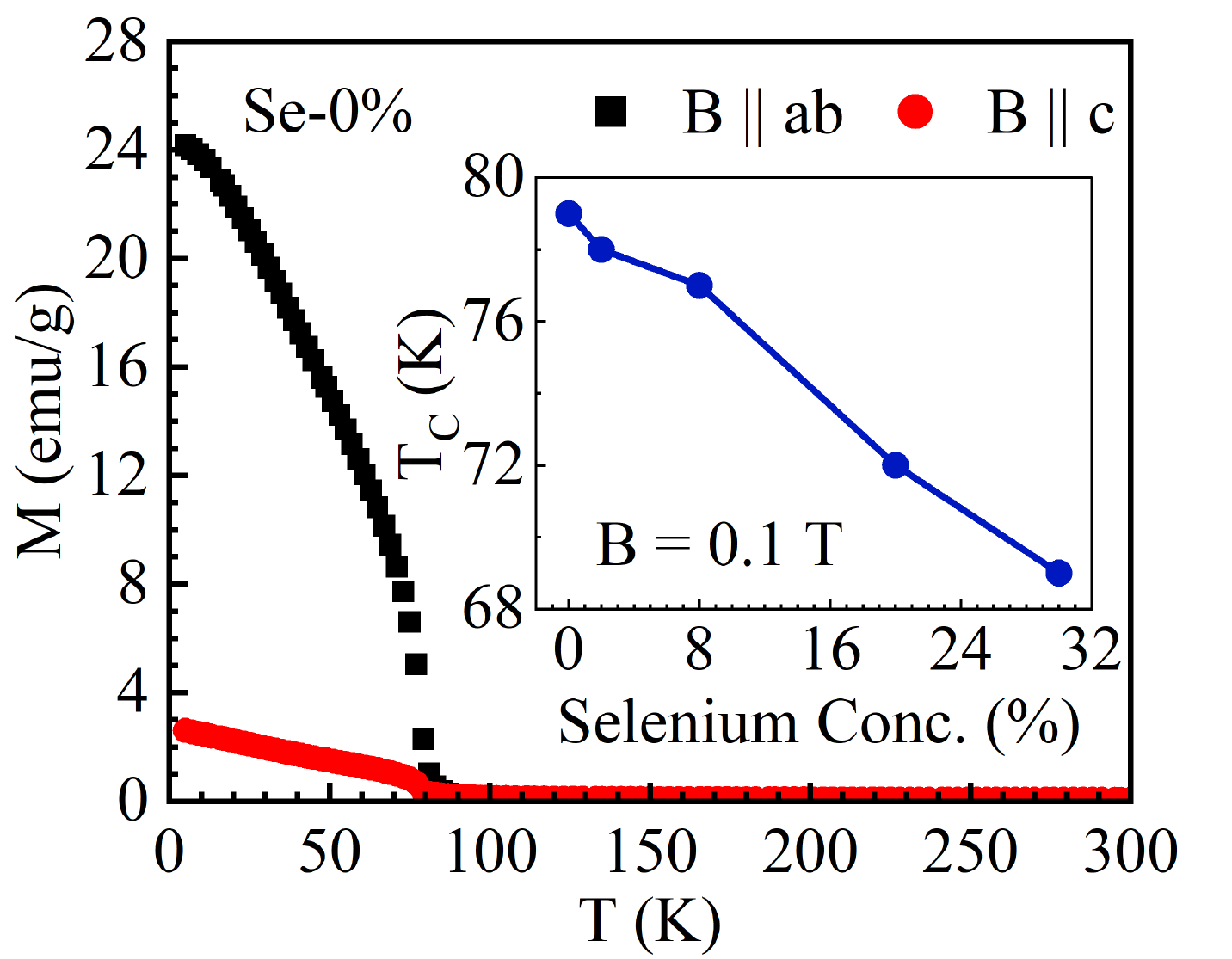}
  \caption{Temperature dependence of magnetization M(T) of Se-0$\%$ single crystal at an applied field of B = 0.1 T with both B $||$ ab and B $||$ c directions. Inset: Variation of Curie temperature T$\mathrm{_C}$ with selenium concentration.}
  \label{fig2}
\end{figure}
\begin{figure*}[]
  \includegraphics[width=0.85\linewidth]{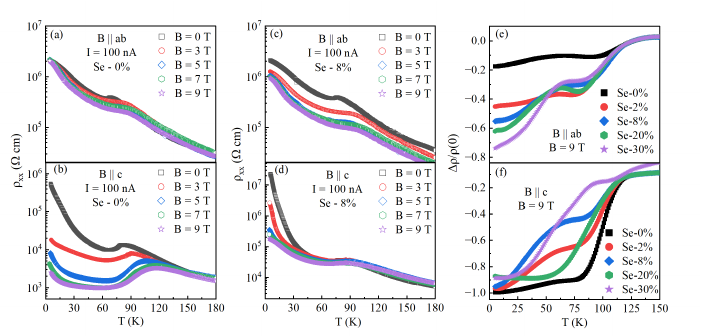}
  \caption{Temperature dependence of longitudinal resistivity $\rho_{\mathrm{xx}}$ of Se-0$\%$ crystal measured at constant magnetic fields with (a) B $||$ ab and (b) B $||$ c. Temperature dependence of longitudinal resistivity $\rho_{\mathrm{xx}}$ of Se-8$\%$ crystal measured at constant magnetic fields with (c) B $||$ ab and (d) B $||$ c. Temperature dependence of magnetoresistivity $\mathrm{\Delta\rho/\rho(0)}$ of Mn$_3$Si$_2$(Te$_{1-x}$Se$_x$)$_6$ single crystals with (e) B $||$ ab and (f) B $||$ c, measured at 9 T.}
  \label{fig3}
\end{figure*}

{\it Results and Discussion:---}
The bulk magnetization data for Se-0$\%$ has been plotted in Fig-\ref{fig2}. It shows that bulk Mn$_3$Si$_2$Te$_6$ single crystal has a Curie temperature T$\mathrm{_C}$ $\sim$ 78 K, which is consistent with previous reports \cite{PhysRevB.103.L161105, Seo2021, Zhang2022, PhysRevB.106.045106, PhysRevB.95.174440, PhysRevB.108.125103, Zhang2024}. The inset of Fig.\ref{fig2} shows the variation of T$\mathrm{_C}$ with selenium concentrations. It is clearly seen that with increasing selenium concentration, the T$\mathrm{_C}$ decreases from 79 K to 69 K as Se substitution level reaches 30$\%$. 

Since the resistivity of Mn$_3$Si$_2$(Te$_{1-x}$Se$_x$)$_6$ varies with different applied currents \cite{Zhang2022}, all electrical transport measurements have been performed at a fixed current of I = 100 nA to avoid any current-induced effect on our observed results. Fig.-\ref{fig3}(a) - (d) shows the temperature dependence of resistivity at different magnetic fields, along the ab-plane and c-axis, for both Se-0$\%$ and Se-8$\%$ samples, respectively. The zero-field (ZF) (with B = 0 T) longitudinal resistivity $\mathrm{\rho_{xx}}$, for both the samples, increases rapidly as we decrease the temperature, which indicates their semiconducting nature. The most remarkable feature of the Se-0$\%$ sample is the CMR along the c-axis, as can be seen in Fig.-\ref{fig3}(b). Here, the resistivity sharply drops by 10$^3$ of magnitude when B $\|$ c. However, from Fig.-\ref{fig3}(a), we can see that such huge drop in resistivity is completely absent when B $\|$ ab. On the other hand, for Se-8$\%$ sample, as shown in Fig.-\ref{fig3}(c), there is a significant drop in $\mathrm{\rho_{xx}}$ even when B $\|$ ab, without significantly affecting the drop along B $\|$ c. To dig further into this contrasting behaviour of $\mathrm{\rho_{xx}}$ for Se-0$\%$ and Se-8$\%$ samples, we perform the magnetoresistivity (MR) measurements of all the samples.   

Fig.-\ref{fig3}(e) and Fig.-\ref{fig3}(f) shows the temperature dependence of MR ($\mathrm{\Delta\rho/\rho(0)}$) for different values of $x$ along the ab-plane and the c-axis, respectively. Here, $\mathrm{\frac{\Delta\rho}{\rho(0)} = \frac{\rho(B) - \rho(0)}{\rho(0)}}$ where $\rho(\mathrm{B})$ is the longitudinal resistivity measured at an applied magnetic field of B, and $\rho(0)$ is the ZF longitudinal resistivity. We find that when B $\|$ ab, for any temperature below 100 K, the value of $\mathrm{\Delta\rho/\rho(0)}$ becomes more and more negative as we increase the selenium concentration. Above 100 K, the MR values along both field directions become negligible due to the absence of magnetic long range order. 
\begin{figure}[htp]
  \includegraphics[width=\linewidth]{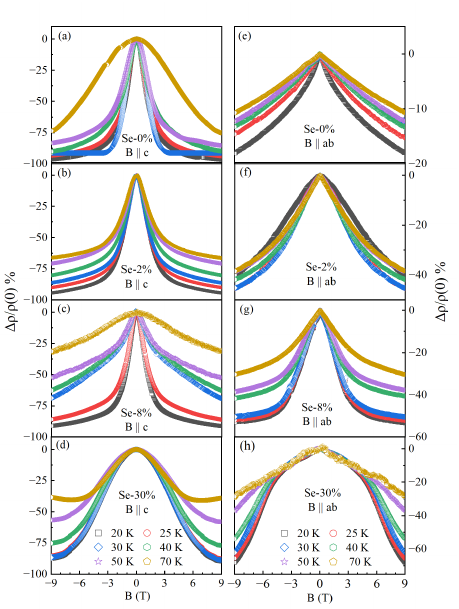}
  \caption{(a) - (d) Magnetic field dependence of MR for Mn$_3$Si$_2$(Te$_{1-x}$Se$_x$)$_6$ at various temperatures with B $\|$ c. (e) - (h) Magnetic field dependence of MR for Mn$_3$Si$_2$(Te$_{1-x}$Se$_x$)$_6$ at various temperatures with B $\|$ ab. The applied bias current is I = 100 nA for all measurements.}
  \label{fig4}
\end{figure}

Next, we perform the MR measurements at constant temperatures for all values of $x$. The MR$\%$ is then calculated and is shown in Fig-\ref{fig4} with B $\|$ c (Fig-\ref{fig4}(a)-(d)), and B $\|$ ab (Fig-\ref{fig4}(e)-(h)), respectively. Here the MR$\%$ is calculated as, $\text{MR}\% = \mathrm{\frac{\Delta\rho}{\rho(0)} \times 100}$. For $x$ = 0, we see that the CMR is present only along the c-axis and gets more pronounced with decreasing temperature. At 20 K, the MR$\%$ reaches almost -98$\%$ when B $\|$ c, as is evident from Fig-\ref{fig4}(a). However, as can be seen in Fig-\ref{fig4}(e), the CMR vanishes when the direction of the magnetic field is changed along the ab-plane. For B $\|$ ab, even at low temperatures of 20 K, the resistance drop is comparatively small $\sim$ -15$\%$. As $x$ is increased, the CMR effect along the c-axis persists for all chemically substituted levels.

The MR curves along the ab-plane are plotted in Fig.-\ref{fig4}(e)-(h) for all $x$. $\neq$ 0. At a low temperature of 20 K, the maximum value of MR is $\sim$ -15$\%$, for $x$ = 0. However, for the Se-2$\%$ substituted samples, the same is $\sim$ -45$\%$, and for Se-8$\%$, Se-20$\%$ and Se-30$\%$ the values drop down to -55$\%$, -60$\%$ and -69$\%$, respectively. Thus, the negative MR increases with the concentration of Se when the field is applied along the ab-plane. This is strikingly different to what happens when the field direction is along the c-axis. The increased magnitude of negative MR for B $\|$ ab in substituted samples is not only limited to lower temperatures but can be seen to persist even at temperatures marginally higher than the transition temperature.
\begin{figure}[htp]
  \includegraphics[width=0.9\linewidth]{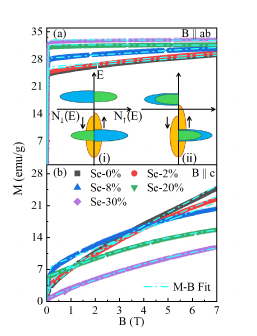}
  \caption{Isothermal magnetization M(B) of Mn$_3$Si$_2$(Te$_{1-x}$Se$_x$)$_6$ single crystals for different concentrations of selenium with (a) B $\|$ ab and (b) B $\|$ c directions, respectively. The M(B) curves are fitted using an empirical model discussed in the text. The dash-dot lines represent the fitted curves. Inset: Schematic illustration of the spin density of states (DOS) for both (i) the unsubstituted and (ii) substituted crystals. Here, blue represents the Mn1 states, green represents the Mn2 states and yellow represents the Te/Se states. }
  \label{fig5}
\end{figure}

To get more insight into this peculiar MR behaviour, the isothermal magnetization measurements were performed on all the samples. Fig.-\ref{fig5} shows the isothermal magnetization curves for different values of $x$ with the field applied along both the ab-plane and c-axis, respectively. It can be seen from Fig.-\ref{fig5}(a) that when the direction of the magnetic field is along the ab-plane, the saturation magnetization M$_s$(at B = 7 T) increases with the increase in selenium concentration. This situation, however, reverses when the field direction is along the c-axis. For B $\|$ c, the magnetization does not saturate at high fields of 7 T owing to the high magnetic anisotropy in the system, which persists even after selenium substitution. Moreover, Fig.-\ref{fig5}(b) clearly shows that the magnetization M(at B = 7 T) along the c-axis decreases as the selenium concentration is increased.

The inset of Fig.\ref{fig5}(a) shows a schematic illustration of the density of states (DOS) for both the unsubstituted and substituted crystals. The blue and green bands refer to the Mn1 and Mn2 states, respectively. The broader yellow bands represent Tellurium or Selenium. The Mn1 and Mn2 spin states in Fig.-\ref{fig5}(a).(i) points opposite to each other, indicating the FIM state. However, in Fig.-\ref{fig5}(a).(ii), they lie in the same direction indicating an FM state \cite{Seo2021}. For a hypothetical FM state, the band gap gets closed and thus, in Fig.-\ref{fig5}(a).(ii), the yellow band is slightly elongated to illustrate this band closing. Now, as is evident from Fig.-\ref{fig5}(a), when the concentration of selenium is increased, the magnetization saturates almost instantly with the applied field. This indicates that, as we increase the concentration of selenium, the ferrimagnetic M(B) curve becomes more ferromagnetic in nature. Comparing this with Fig.-\ref{fig5}(a).(ii) we conclude that when B $\|$ ab, the band gap closes as we increase the selenium concentration. However, the magnetization behaviour along the c-axis is opposite to what we observe along the ab-plane.

This contrasting magnetization behaviour along different directions of the applied magnetic field can be attributed to the COC developed in Mn$_3$Si$_2$(Te$_{1-x}$Se$_x$)$_6$ crystals due to the Te atoms. It is well known that the COC that circulates along the Te edges produces a magnetic moment $\mathrm{M_{COC}}$ of its own which orients along the c-axis as the orbital currents flow along the ab-plane \cite{Zhang2022}. Thus, the net magnetic moment is a combination of both M$\mathrm{_{COC}}$ and the moments arising from the Mn atoms, M$\mathrm{_{Mn}}$. For unsubstituted Mn$_3$Si$_2$Te$_6$ crystal, the Mn atoms (M$\mathrm{_{Mn}}$) predominantly point along the ab-plane unlike M$\mathrm{_{COC}}$ which points along the c-axis. The coupling between the M$\mathrm{_{COC}}$ and M$\mathrm{_{Mn}}$ gives rise to a unique SOC effect, which on the application of magnetic field underpins the CMR in the Se-0$\%$ sample \cite{Seo2021, Zhang2022}. Thus, the saturation magnetization and CMR are interlinked with each other. Substituting the tellurium site of Mn$_3$Si$_2$Te$_6$ with selenium decreases the orbital currents and its corresponding domain area decreases as selenium is smaller in size than tellurium. Thus, the orbital moment contribution $\mathrm{M_{COC}}$ arising due to the COC of tellurium become smaller with increasing selenium substitution. So the coupling between M$\mathrm{_{COC}}$ and M$\mathrm{_{Mn}}$ becomes weaker with increasing concentration of selenium. Thus, the net magnetic moment of the substituted Mn$_3$Si$_2$(Te$_{1-x}$Se$_x$)$_6$ crystals is dominated by the Mn atoms in contrast to the unsubstituted crystal. Due to this decreased coupling between M$\mathrm{_{COC}}$ and M$\mathrm{_{Mn}}$, the SOC developed due to the combined effect of both gets reduced, leading to the MR and the magnetization behaviour of the substituted crystals as observed in Fig.-\ref{fig4} and Fig.-\ref{fig5}.

\begin{figure}[htp]
  \includegraphics[width=0.95\linewidth]{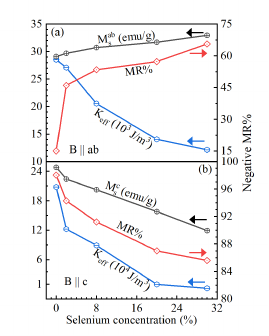}
  \caption{Saturation magnetization M$_{\mathrm{s}}$ (at B = 7 T), effective anisotropy constant K$\mathrm{_{eff}}$ and maximum negative MR$\%$ (at B = 9 T) as a function of selenium concentrations $x$ with (a) B $\|$ ab and (b) B $\|$ c directions respectively. The saturation magnetization and effective anisotropy constant are plotted along the left y-axis with black and blue symbols, respectively. On the right y-axis, the negative MR$\%$ is plotted with red symbols. The continuous lines act as a guide to the eye.}
  \label{fig6}
\end{figure}

Fig.-\ref{fig6} simultaneously shows the saturation magnetization M$\mathrm{_s}$, effective anisotropy constant K$\mathrm{_{eff}}$ and maximum negative MR as a function of selenium concentration $x$ in both in-plane (IP) and out-of-plane (OOP) configurations, respectively. The K$\mathrm{_{eff}}$ has been calculated by fitting the isothermal magnetization data in Fig-\ref{fig5} with an empirical model for the M-B curve for different $x$ values along both field directions. The M(B) measurements for all the samples were recorded using a five-quadrant measurement. The curves show almost negligible hysteresis, and thus, only the positive cycle of the M(B) curves are shown in Fig.-\ref{fig5}. Our analysis shows that COC plays a crucial role in governing the magnetization and MR behaviour of all the samples. Hence, to find the causal relation between COC and selenium substitution, we need to get an idea of the strength of the SOC for all the samples. Qualitatively, the behaviour of K$\mathrm{_{eff}}$ is analogous to the behaviour of SOC. Thus, calculating the variation of K$\mathrm{_{eff}}$ with the selenium concentration will give us a clear picture of the behaviour of SOC with the same. The following analysis is based on the Stoner-Wolfarth model and the Law of Approach to Saturation  \cite{las1,las2,las3,las4,las5}. 

According to the Stoner-Wolfarth model, the total energy E of a ferromagnet is given by:
\begin{align}
    \mathrm{E} = \mathrm{K_{eff}V\sin^2\theta - M_s BV\cos(\theta-\phi)} \label{SW}
\end{align}
where $\mathrm{K_{eff}}$ = effective anisotropy constant, V = volume of the particle, M$\mathrm{_s}$ =  saturation magnetization, B = applied magnetic field, $\theta$ = angle of magnetization with respect to the easy axis, and $\phi$ = angle of the applied field with respect to the easy axis. Based on this model, the descending branch of M(B) curve at high fields can be approximated by:
\begin{align}
    \mathrm{|M|} = \mathrm{M_s\Bigg[1 - \frac{H_K^2}{C\cdot B^2}\Bigg]} \label{LAS}
\end{align}
where C is a constant depending upon the system, and $\mathrm{H_K}$ = 2$\mathrm{K_{eff}}$/$\mathrm{M_s}$. This is called the Law of Approach to Saturation \cite{las3,las5}. Based on the above two equations, the expression for fitting the entire M(B) curve is given by \cite{las5}:
\begin{align}
    \mathrm{\frac{M}{M_s}} = \mathrm{g(B)f(\xi, 0) + (1 - g)f(\xi,\infty)} \label{M(B) fit}
\end{align}
where,
\begin{eqnarray}
    \mathrm{f(\xi,0)} &=& \mathrm{\coth\Bigg(\frac{BM_s}{k_B T}\Bigg) - \frac{k_B T}{BM_s}} \\
    \mathrm{f(\xi,\infty)} &=& \mathrm{a_1\tanh\Bigg(b_1\frac{BM_s}{k_B T}\Bigg)+a_2\tanh\Bigg(b_2\frac{BM_s}{k_B T}\Bigg)}\nonumber \\ 
    &&+\mathrm{a_3\tanh\Bigg(b_3\frac{BM_s}{k_B T}\Bigg)} \\
    \mathrm{g(B)} &=& \mathrm{c_1\tanh\Bigg(d_1\frac{B}{H_K}\Bigg)+c_2\tanh\Bigg(d_2\frac{B}{H_K}\Bigg)}\nonumber \\ 
    &&+\mathrm{c_3\tanh\Bigg(d_3\frac{B}{H_K}\Bigg)}
\end{eqnarray}
Here, a$_\mathrm{i}$, b$_\mathrm{i}$, c$_\mathrm{i}$ and d$_\mathrm{i}$ are all constants with i = 1,2, and 3, k$\mathrm{_B}$ is the Boltzmann constant and T is temperature. We use Eqn.-\eqref{M(B) fit} for fitting the M(B) curves along both field directions.

It can be seen from Fig.-\ref{fig6}(a) that on increasing selenium concentration $x$, the saturation magnetization M$^{\mathrm{ab}}_{\mathrm{s}}$ (along ab-plane) increases and so does the negative MR$\%$. However, the effective anisotropy constant K$\mathrm{_{eff}}$ reduces with increasing selenium concentrations. Qualitatively, a decrease in the effective anisotropy indicates a decrease in the overall SOC. Thus, we can deduce that the SOC induced due to the heavy Te atoms gets reduced when it is substituted with lighter Se atoms. Fig-\ref{fig6}(a) shows a correlation between the saturation magnetization and the MR response, indicating that one behaviour is influenced by the other. It is evident that selenium substitution leads to a higher value of saturation magnetization M$^{\mathrm{ab}}_{\mathrm{s}}$. Since the observed magnetization is a combined effect of $\mathrm{M_{COC}}$ and M$\mathrm{_{Mn}}$, a higher magnetization value indicates a weaker cancellation effect between the $\mathrm{M_{COC}}$ and M$\mathrm{_{Mn}}$.

When the field is along the ab-plane, the Mn moments readily point along the applied field as it is the easy plane. For the Se-0$\%$ sample, the $\mathrm{M_{COC}}$ points along the c-axis only, as the COC predominantly circulates along the ab-plane itself. The generated COC domains are much smaller and decrease further with an increase in the magnetic field. For the Se-0$\%$ sample, the magnetization does not readily saturate at smaller field values since the COC domains do not completely vanish. Thus, the effective magnetic moment along the ab-plane is somewhat less than what it should have been if there had been no $\mathrm{M_{COC}}$ along the c-axis. As we substitute tellurium with selenium, the small COC domains shrink. The effective moment increases with the increasing concentration of selenium, and the FIM state eventually becomes FM, as can be seen from Fig.-\ref{fig5}(a). The reduction in $\mathrm{M_{COC}}$ also decreases the SOC along the ab-plane, as is evident from Fig.-\ref{fig6}(a). With the decreasing SOC and the strengthening of FM character, the DOS of the substituted samples closes the band gap as is illustrated in Fig.-\ref{fig5}(a).(ii). Hence, more electrons become available for conduction for the substituted samples. Thus, we can see a simultaneous increase in both effective moment and negative MR along the ab-plane with increasing selenium substitution.

For B $\|$ c, all the three curves of M$^{\mathrm{c}}_{\mathrm{s}}$, negative MR$\%$ and K$\mathrm{_{eff}}$ follow similar behaviour as we increase the selenium concentration which can be seen in Fig.-\ref{fig6}(b). The striking similarity between the three curves in Fig.-\ref{fig6}(b) again ascertains that with an increase in selenium concentration, the $\mathrm{M_{COC}}$ gradually decreases, which decreases the effective moment along the c-axis. The CMR also follows the same trend.  

When the field is along the c-axis, depending on the field value (whether negative or positive), the COC circulates in either a clockwise or anticlockwise direction, respectively. As the field is increased, the domain of a particular chirality is favoured, and it enlarges in size, which causes the moments of the same chirality, $\mathrm{M_{COC}}$, to align collectively towards the field direction, thereby increasing the COC moments. Since the Mn atoms do not readily point along the c-axis, the M$\mathrm{_{Mn}}$  aligns along the field direction slowly with increasing field value. This leads to a coupling between $\mathrm{M_{COC}}$ and M$\mathrm{_{Mn}}$ that causes a sharp decrease in electron scattering \cite{Zhang2022} and thus promotes the negative CMR along the c-axis for Se-0$\%$ crystal. However, when selenium concentration is increased, the expansion of a single COC domain is impeded as selenium does not produce any COC. This leads to a reduction in the size of individual COC domains and decreases the COC moments $\mathrm{M_{COC}}$. Thus, the sharp reduction in electron scattering is partially suppressed with increasing substitution, which leads to a decrease in the CMR effect but not its complete disappearance as the COC domains of the remaining Te atoms still cause sufficient reduction in electron scattering which paves the way for CMR. Thus the contrasting behaviour of M$_{\mathrm{s}}$ and negative CMR on changing the direction of the applied field in Fig.-\ref{fig6} points to the fact that the entire scenario is governed by COC moments, which has opposite effects along the two crystallographic axes. 

{\it Conclusion:---}
We have studied the magnetization and electrical transport properties of FIM nodal-line semiconductor Mn$_3$Si$_2$(Te$_{1-x}$Se$_x$)$_6$ as a function of substitution concentrations. We observe that as we substitute the tellurium (Te) with selenium (Se), the COC domains decrease in size. This leads to a decrease in the moment induced by the COC ($\mathrm{M_{COC}}$) and subsequently coupling between $\mathrm{M_{COC}}$ and M$\mathrm{_{Mn}}$, which has opposite effects along the ab-plane and c-axis, respectively. The ab-plane magnetization, as well as the negative MR, increases as we increase the concentration of selenium, and the isothermal magnetization curves of the selenium substituted samples show a much more FM character than the FIM unsubstituted sample. On the other hand, the c-axis magnetization and the CMR tends to decrease, albeit marginally, with increasing chemical substitution. Thus, our findings reveal the close relationship between magnetization and CMR. By chemical substitution of tellurium of Mn$_3$Si$_2$Te$_6$, we are able to increase the negative MR along the ab-plane without hindering much the CMR effect along the c-axis, which could be potentially useful for applications.

{\it Acknowledgements:---}
The authors acknowledge IIT Kanpur and the Department of Science and Technology, India, [Order No. DST/NM/TUE/QM-06/2019 (G)] for financial support. A.D. thanks PMRF for financial support.

\end{document}